\documentclass[onecolumn,preprint,aps,pra,showpacs]{revtex4-2}%
\pdfoutput=1
\usepackage{amsmath}
\usepackage{amsfonts}%
\usepackage{amssymb}%
\usepackage[pdftex]{graphicx}
\providecommand{\U}[1]{\protect\rule{.1in}{.1in}}

\begin{document}
\title{Modification of Selection Rules of Landau-Quantized Electron by Coupling with
Obliquely Irradiated Optical Vortex Beams}
\author{Hirohisa T. Takahashi}
\affiliation{Academic Support Center, Kogakuin University of Technology \& Engineering,
2665-1 Nakano, Hachioji, Tokyo 192-0015, Japan}
\email{kt13673@ns.kogakuin.ac.jp}
\author{Jun-ichiro Kishine}
\affiliation{Division of Natural and Environmental Sciences, The Open University of Japan,
Chiba 261-8586, Japan}
\affiliation{Institute for Molecular Science, 38 Nishigo-Naka, Myodaiji, Okazaki 444-8585, Japan}

\begin{abstract}
We discuss selection rules and intensities of photocurrent of a
two-dimensional electron gas in a strong magnetic field via absorptions of
orbital angular momentum carried by obliquely irradiated optical vortex beams.
By angular deflection, optical vortex beams on the two-dimensional electron
gas interface can be seen as a superposition of the various orbital angular
momentum states. As a result, it is demonstrated that it is yielded that the
angular momentum selection rules and induced currents are modified\ from the
case of vertical incidence.

\end{abstract}
\volumeyear{year}
\volumenumber{number}
\issuenumber{number}
\eid{identifier}
\date[Date text]{date}
\startpage{1}
\endpage{102}
\maketitle

\section{Introduction}

It is theoretically pointed out that lights with a circular polarization can
exert torque by Poynting in 1909,\cite{Poynting1909} and it is experimentally
demonstrated that circularly-polarized lights carry angular momentum of
$\pm\hbar$, namely, spin angular momentum (SAM) by Beth in
1936.\cite{Beth1936} In 1992, an optical vortex beam (OV beam) was proposed by
L. Allen \textit{et al.}. They pointed out that light beams can also carry an
intrinsic orbital angular momentum (OAM).\cite{Allen1992} OV carries an OAM
$\ell\hbar$ by an azimuthal phase dependence of $e^{i\ell\phi}$ with an
integer $\ell$. The integer $\ell$ is the winding number of the helical
wavefront.\ The azimuthal phase dependence makes $\ell$ intertwined helical
wavefront. Hereafter, we refer to the winding number $\ell$ as an orbital
number. Then, it is known that OV with non-zero OAM has a phase singularity on
its optical axis. A Laguerre-Gaussian mode (LG-mode) is well known as one of
the radial\ intensity modes. Several ways to product the LG-mode OV are known:
directly productions,\cite{Tamm1990,Harris1994} conversion from
Hermite-Gaussian modes,\cite{Allen1992,Beijersbergen1993} the use of spiral
phase plates,\cite{Beijersbergen1994} the use of computer-generated
holograms,\cite{Heckenberg1992} diffractive optics,\cite{Kennedy2002} or
semiconductor microcavity.\cite{Nakagawa2020} The LG-mode beam has the nature
of gradually expanding the beam as it propagates.

Besides the LG-mode, Bessel-mode OV is also well known as the radial function
mode.\cite{Jentschura2011} In general, the amplitude of Bessel-mode OV in
cylindrical coordinates $(r,\phi,z)$ can be described by
\begin{equation}
E(r,\phi,z)\propto J_{\ell}(k_{\perp}r)e^{ik_{z}z}e^{i\ell\phi},
\end{equation}
where $J_{\ell}(x)$ is an $\ell$-th order Bessel function of the first kind,
$k_{z}$ and $k_{\perp}$ are respectively the longitudinal and transverse
wavenumber satisfying $k=\sqrt{k_{z}^{2}+k_{\perp}^{2}}$. It thus has a
cylindrical intensity distribution consisting of bright and dark rings.\ In
particular, since\ its intensity ($I\propto\left\vert E\right\vert ^{2}$),
obeys $I(r,\phi,z\geq0)=I(r,\phi)$, the intensity is invariant under the
propagation of the beam. That is to say, the Bessel-mode beam can be
considered diffraction-free.\cite{Durnin1987a,McGloin2005} The Bessel-mode OV
can be created, in the back focal plane of a convergent lens by a plane
wave,\cite{Durnin1987b} or in that of an axicon from a Gaussian
beam.\cite{Indebetouw1989} Besides, it can be created\ by the use of
computer-generated holograms,\cite{Vasara1989} or by a Fabry-Perot
resonator.\cite{Cox1992}

The transfer of optical OAM to physical matters has much got attracted and
actively\ been researched. For examples, the transfer to classical particles
as an optical tweezer,\cite{Amos1995,Friese1995} to exciton center-of-mass
motion,\cite{Shigematsu2016} to the bounded electron in
atoms,\cite{Schmiegelow2016} or the laser ablation
technique\cite{Hamazaki2010} have been demonstrated.\ Various theoretical
predictions in condensed matter physics have also done: the optical absorption
by semiconductors,\cite{Quinteiro2009a} an electric current density in a
semiconducting stripe,\cite{Watzel2012} the excitation of multipole plasmons
in metal nanodisks,\cite{Sakai2015}, the spin and charge transport on surface
of topological insulators,\cite{Shintani2016}, the generation of skyrmionic
defects in chiral magnets,\cite{FujitaSato2017} and the creation of
superconducting vortices,\cite{Yokoyama2020} among other things.

However, it is known that an exchange of the optical OAM does not occur in an
electric dipole transition in atoms and molecules.\cite{Babiker2002} It is
still an open question how optical OAM affects electronic transitions. To
fully exchange the optical OAM to the electron center-of-mass motion in
electric dipole transitions, it is needed the conformity of the electron
distribution with the OV intensity distribution. Therefore, in our previous
paper,\cite{Takahashi2018} we considered\ two-dimensional electron gas (2DEG)
as one of the good candidates. It is because an axial symmetric 2DEG has the
cylindrical electron distribution as the coherent state. Thus our previous
work was focused on the case of vertical incidence of OV beams.\ As a result,
we obtained the selection rules in the electric dipole transitions from the
lowest Landau level (LLL) to the second Landau level (2LL) as$\ $%
\begin{equation}
\ell=2,\sigma=-1\text{, and }\ell=0,\sigma=1\label{Selection_rule0}%
\end{equation}
with optical spin number $\sigma$. However, from experimental viewpoints, it
is quite natural to consider the case of oblique incidence. In the case of
oblique incidence, the incident beam carries a definite OAM, but the
irradiated beam consists of a superposition of various OAM
states.\cite{Vasnetsov2005,Lavery2011} Therefore, it is expected that the AM
selection rules and behaviors of OV-induced photocurrents are modified. In
this paper, we investigate the modification of the selection rules from Eq.
(\ref{Selection_rule0})\ and numerically demonstrate the incident angle and
magnetic field dependence of OV-induced photocurrents by\ the oblique
irradiation (angular deflection) of OV beams to the reference axis of 2DEG.

This paper is organized as follows. In Section II, we describe a
circularly-polarized Bessel-mode optical vortex (CPBOV) beam in oblique
irradiation. We calculate the induced photocurrent in 2DEG by obliquely
irradiated CPBOV in Sec. III. In Sec. IV, we show numerical results of
the\ incident angle and magnetic field dependence of the OV-induced
photocurrent. Sec. V is reserved for conclusions and remarks.

\section{Description of Obliquely Incident Optical Vortex Beam}

It is crucial for studying the properties of light to separate a total angular
momentum (TAM) into a spin and orbital AM since they are then separately
conserved when interacting with particles. In electromagnetism, the TAM
density of the electromagnetic field is generally given by the vector product
of the radial vector and the linear momentum density. But such a form is
typically similar to the form for an orbital angular momentum for a rigid body
in mechanics. This curious fact shows the difficulties in separating the TAM
into the spin and orbital parts for an electromagnetic field. In a paraxial
approximation, such the light beam has a well-defined separation of the TAM
into the spin and orbital AMs. Then it becomes natural to define optical OAM
to its optical axis and optical OAM is identified as the eigenvalue of the
quantum mechanical AM operator.\cite{vanEnk1992} Therefore, a discrete OAM
spectrum (orbital number) is produced. This is the result of an axially
symmetric structure of the light beam.

Generally, the optical OAM depends on the choice of the reference axis of the
system.\cite{Molina-Terriza2002,ONeil2002} If the axial symmetric structure is
broken, \textit{e.g.} by a\ lateral displacement or angular tilt of the beam,
the reference axis is changed and it is impossible to determine the reference
axis for a nonsymmetric beam. In particular, the initially carrying OAM
transforms to the superposition of OAM states in a changed coordinate frame.
Nevertheless, such the transformed beam with a well-defined OAM\ can be
described in the form of an infinite set of azimuthal harmonics in the form of
Bessel functions.\cite{Vasnetsov2005} Therefore, it is allowed the evaluation
of the interaction of the nonsymmetric beam with microscopic particles in a
quantitative way. In this paper, to investigate contributions of an OAM of
light to 2DEG via AM selection rules, we adopt an OV beam in a paraxial
approximation, in particular, a circularly polarized Bessel-mode optical
vortex beam (CPBOV beam). We then concentrate on the case of the angular
deflection to the reference axis and describe its vector potential of CPBOV in
this section.

We assume the vector potential of OV is taken in the form of the monochromatic
beam, $\mathbf{A}^{\text{OV}}\left(  \mathbf{r},t\right)  =\mathbf{A}%
^{\text{OV}}\left(  \mathbf{r}\right)  \mathrm{e}^{-i\omega t}$.\ By
introducing the longitudinal wavenumber along to the beam axis $k_{\parallel}$
and its transverse wavenumber $k_{\perp}$, the spatial part of the vector
potential of CPBOV traveling along $Z$-direction\ in the beam flame is given
by\cite{Jentschura2011,Matula2013}%

\begin{equation}
\mathbf{A}_{\ell,\sigma}^{\text{OV}}(\mathbf{r|}\theta)=\mathbf{\varepsilon
}_{\sigma}A_{0}\sqrt{\frac{k_{\perp}}{2\pi}}iJ_{\ell}(k_{\perp}R_{\perp
})\mathrm{e}^{i\ell\Phi}\mathrm{e}^{ik_{\parallel}Z}%
,\label{Tilted_vector_potential_pm}%
\end{equation}
where $A_{0}$ is a real constant, $\ell$\ is a topological number of OAM, and
$J_{\ell}(x)$ is $\ell$-th order Bessel function of the first kind. Because of
$J_{\ell}(k_{\perp}R_{\perp})$, the beam profile has cylindrical intensity
distribution. The superscript 'BF' indicates the quantity in the beam flame.
$k=\omega/c=\sqrt{k_{\perp}^{2}+k_{\parallel}^{2}}$ is the magnitude of
wavenumber with the speed of light in vacuum $c$, and the frequency $\omega$.
By introducing a ratio $\alpha=k_{\perp}/k_{z}(\ll1)$, the transverse
wavenumber is given by $k_{\perp}=\sqrt{k^{2}-k_{\parallel}^{2}}\sim\alpha k$.
The polarization vectors in the beam flame are taken on the helicity basis,
\begin{equation}
\mathbf{\varepsilon}_{\sigma}=\frac{1}{\sqrt{2}}\left(  \mathbf{e}_{X}%
+i\sigma\mathbf{e}_{y}\right)  ,\ \text{for }\sigma=\pm1,\label{Cir_pol}%
\end{equation}
which gives $\hat{S}_{Z}\mathbf{A}_{\ell,\sigma}^{\text{OV}}\left(
\mathbf{r}\right)  =\hbar\sigma\mathbf{A}_{\ell,\sigma}^{\text{OV}}\left(
\mathbf{r}\right)  $. Here $\hat{S}_{Z}$ is the $Z$-component of the SAM
operator. The $Z$-component of the OAM operator is%
\begin{equation}
\hat{L}_{Z}=-i\hbar\frac{\partial}{\partial\Phi},
\end{equation}
which gives
\begin{equation}
\hat{L}_{Z}\mathbf{A}_{\ell,\sigma}^{\text{OV}}\left(  \mathbf{r}\right)
=\hbar\ell\mathbf{A}_{\ell,\sigma}^{\text{OV}}\left(  \mathbf{r}\right)  .
\end{equation}
Then, the $Z$-component of\ the TAM of the beam, $\hat{J}_{Z}=\hat{L}_{Z}%
+\hat{S}_{Z}$, is given by $J=\ell+\sigma$.

\begin{figure}
[ptb]
\begin{center}
\includegraphics[
height=2.8072in,
width=4.7816in
]%
{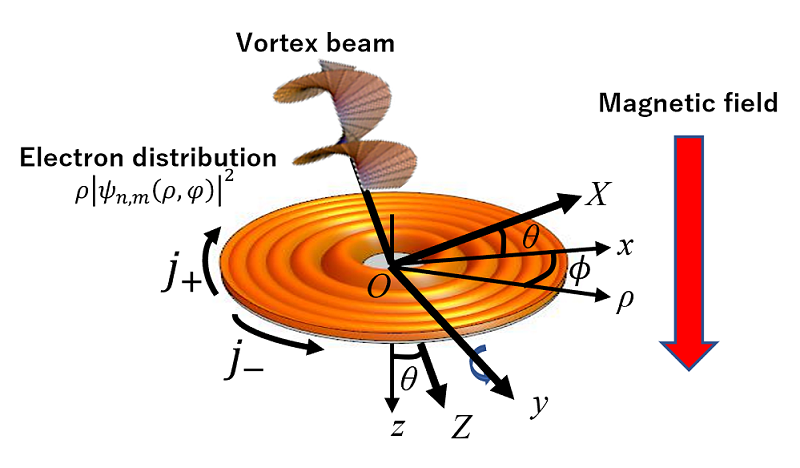}%
\caption{The setup configuration. The laboratory flame $(x,y,z)$ is chosen the
$xy$-plane being on the 2DEG plane and the $z$-axis being on the normal to the
plane. The $z$-axis is then the reference axis. The azimuthal angle $\phi$ of
the cylindrical coordinates on the laboratory frame is measured from the
$x$-axis. The beam flame $(X,y,Z)$ is chosen as the $Z$-axis to the beam axis
and the $Xy$-plane is rotated $xy$-one\ around the $y$-axis with angle
$\theta$.}%
\label{SetupConfig}%
\end{center}
\end{figure}

To discuss the oblique irradiation of OVs to a 2D system after this, since the
expression (\ref{Tilted_vector_potential_pm}) is described in the beam
flame,\ we need to describe the vector potential on the 2D plane (the
Laboratory flame). We then describe the beam frame by the Cartesian coordinate
as $(X,y,Z)$, where the light beam is traveling along $Z$-axis. $R_{\perp}$
and $\Phi$ are the radial component and azimuthal angle on $Xy$-plane,
respectively. We here refer to the rotation around the $y$-axis\ by the angle
$\theta$\ from the laboratory frame $(x,y,z)$ to the beam frame $(X,y,Z)$\ as
shown in Fig.\ref{SetupConfig}. Thus we can consider the beam traveling\ at an
incident angle $\theta$ to the reference axis $z$. The unit vectors
$\mathbf{e}_{i}$ are then transformed by the rotation matrix $\mathcal{R}%
_{y}(\theta)$ as
\begin{equation}%
\begin{pmatrix}
\mathbf{e}_{X}\\
\mathbf{e}_{y}\\
\mathbf{e}_{Z}%
\end{pmatrix}
=\mathcal{R}_{y}^{-1}\left(  \theta\right)
\begin{pmatrix}
\mathbf{e}_{x}\\
\mathbf{e}_{y}\\
\mathbf{e}_{z}%
\end{pmatrix}
\text{ with }\mathcal{R}_{y}\left(  \theta\right)  =%
\begin{pmatrix}
\cos\theta & 0 & -\sin\theta\\
0 & 1 & 0\\
\sin\theta & 0 & \cos\theta
\end{pmatrix}
.\label{Rotation_matrix}%
\end{equation}%
First, we discuss the transformation of the polarization. By the
transformation (\ref{Rotation_matrix}),\ the polarization in the Laboratory
frame $\mathbf{\eta}_{\sigma}$\ is given by%
\begin{align}
\mathbf{\eta}_{\sigma}  & =\mathcal{R}_{y}\left(  \theta\right)
\mathbf{\varepsilon}_{\sigma}\nonumber\\
& =\frac{1}{\sqrt{2}}\left(  \cos\theta\boldsymbol{e}_{x}+i\sigma
\boldsymbol{e}_{y}-\sin\theta\boldsymbol{e}_{z}\right)  .
\end{align}
Thus, in the laboratory frame, the polarization of the obliquely-irradiated
beam with circular polarization behaves as the elliptic polarization with
ellipticity, $\tan\chi=1/\cos\theta$. Also, the circular shape of the profile
of CPBOV changes into an elliptic shape. Because the argument of $J_{\ell
}(k_{\perp}R_{\perp})$ in the Laboratory flame is written by%
\begin{equation}
k_{\perp}R_{\perp}=k_{\perp}\rho\left[  \cos^{2}\theta\cos^{2}\phi+\sin
^{2}\phi\right]  ^{\frac{1}{2}},\label{Bessel_factor}%
\end{equation}
where we introduced the cylindrical coordinates in the laboratory frame as
$\left(  \rho,\phi,z\right)  $ as shown in Fig.\ref{SetupConfig}. Furthermore,
to fully describe the vector potential of obliquely irradiated OV in
Laboratory flame, we also need the transformation of the azimuthal phase
factor $\mathrm{e}^{i\ell\Phi}$ and the traveling wave sector $\mathrm{e}%
^{ik_{\parallel}Z}$. The azimuthal phase factor $\mathrm{e}^{i\ell\Phi}$ are
rewritten as%
\begin{align}
\mathrm{e}^{i\ell\Phi}  & =\mathrm{e}^{i\ell\arctan\left(  \frac{\tan\phi
}{\cos\theta}\right)  }\nonumber\\
& =\frac{\left(  \cos\theta\cos\phi+i\sin\phi\right)  ^{\ell}}{\left(
\cos^{2}\theta\cos^{2}\phi+\sin^{2}\phi\right)  ^{\frac{\ell}{2}}%
}.\label{Azimuthal_factor}%
\end{align}
And the traveling wave sector $\mathrm{e}^{ik_{\parallel}Z}$\ can be expanded
as%
\begin{equation}
\mathrm{e}^{ik_{\parallel}Z}=\mathrm{e}^{ik_{\parallel}z\cos\theta}%
{\sum\limits_{L=-\infty}^{\infty}}\mathrm{e}^{iL\frac{\pi}{2}}J_{L}%
(k_{\parallel}\rho\sin\theta)\mathrm{e}^{iL\phi},\label{MultipoleFactor}%
\end{equation}
by the Jacobi-Anger expansion (the partial wave expansion) and the Bessel
function of the first kind with $L$ in an integer.\cite{Abramowitz1965}\ When
we restrict our discussion to the region that the incident angle $\theta$ is
sufficiently small.\ Then, it satisfies $\sin\theta\sim\theta$ and $\cos
\theta\sim1$. Thus the vector potential in the laboratory flame arrives at%
\begin{align}
\mathbf{A}_{\ell,\sigma}^{\text{OV}}(\boldsymbol{r}|\theta)  & \sim\left(
\boldsymbol{e}_{x}+i\sigma\boldsymbol{e}_{y}-\theta\boldsymbol{e}_{z}\right)
A_{0}\sqrt{\frac{k_{\perp}}{4\pi}}J_{\ell}(k_{\perp}\rho)\nonumber\\
& \times\mathrm{e}^{ik_{\parallel}z}{\sum\limits_{L=-\infty}^{\infty}}%
J_{L}(k_{\parallel}\rho\text{ }\theta)\mathrm{e}^{i\left(  \ell+L\right)
\phi}\mathrm{e}^{i\left(  L+1\right)  \frac{\pi}{2}}%
.\label{Tilted_vector_potential_lab2}%
\end{align}
It appears the changes of the azimuthal phase factor as $\mathrm{\exp}\left[
i\left(  \ell+L\right)  \phi\right]  $ and of the radial profile function as
$J_{\ell}(k_{\perp}\rho)J_{L}(k_{\parallel}\rho$ $\theta)$. This physical
meaning is that CPBOVs can be seen as a superposition of waves with OAM,
$\ell+L$.\cite{Cohen-Tannoudji1997,Vasnetsov2005}

As mentioned in detail later, the partial wave expansion
(\ref{Tilted_vector_potential_lab2}) plays an important role and gives birth
to curious results. As the CPBOVs are constituted by\ the superposition of the
OAM $\ell+L$ state, they modify the optical AM selection rules. Also, the
factor $J_{L}(k_{\parallel}\rho$ $\theta)$\ also contributes to the
oscillatory behavior of the OV-induced photocurrent in the change of incident
angle $\theta$.

\section{Photocurrent Induced by Obliquely Irradiated Optical Vortex Beam}

Landau-quantized energy spectrum appears when a strong magnetic field is
uniformly applied to a 2-dimensional electron gas (2DEG) perpendicular to the
specimen surface. The energy spectrum of the standard Landau-quantized state,
$E_{N}=\hbar\omega_{c}\left(  N+1/2\right)  $, appears under a Landau gauge
(suitable for a\ rectangle specimen), where $N$\ is a positive integer,
$\omega_{c}=eB/m_{e}$ is the cyclotron frequency with the elementary charge
$e\left(  >0\right)  $, the strength of the magnetic field $B $, and the
electron mass $m_{e}$. For a\ circular disk geometrical specimen, the energy
spectrum explicitly depends on an OAM of the electron under the symmetric
gauge. This is a manifestation that the symmetric gauge preserves axial
symmetry. Thus an electron OAM becomes a good quantum number. For these
reasons, a circular disk geometrical specimen is the best choice to observe
the induced current by the interaction with lights carrying an OAM.

The Hamiltonian for 2DEG applying the perpendicular magnetic field in the
$z$-axis direction is given by%

\begin{equation}
H_{0}=\frac{1}{2m_{e}}\left[  -i\hbar\mathbf{\nabla}+e\mathbf{A}^{\text{ext}%
}(\mathbf{r})\right]  ^{2},\label{Non-perturmativeHamil}%
\end{equation}
with symmetric gauge, $\mathbf{A}^{\text{ext}}(\mathbf{r})=\left(
-By/2,Bx/2,0\right)  $. Then the energy spectrum and the wavefunction in
cylindrical coordinates $\left(  \rho,\phi,z\right)  $\ in the laboratory
flame are respectively written as%
\begin{align}
E_{n,m} &  =\hbar\omega_{c}\left(  N+\frac{1}{2}\right) \nonumber\\
N &  =n+\frac{|m|+m}{2},\nonumber\\
n &  =0,1,2,...,\text{and }m=0,\pm1,\pm2,...,\label{Energy_Landau}%
\end{align}
and,%
\begin{align}
\Psi_{nm}\left(  \rho,\phi\right)   & =N_{nm}\mathrm{e}^{-\frac{\rho^{2}%
}{4l_{B}^{2}}}\left(  \frac{\rho}{l_{B}}\right)  ^{|m|}L_{n}^{|m|}\left(
\frac{\rho^{2}}{2l_{B}^{2}}\right)  \frac{\mathrm{e}^{im\phi}}{\sqrt{2\pi}%
}\nonumber\label{WF_of_electron}\\
& \equiv\mathfrak{R}_{n,m}(\rho)\frac{\mathrm{e}^{im\phi}}{\sqrt{2\pi}},
\end{align}
where $n$ is a principal quantum number and $m$, an electron orbital
number.\cite{LaudauLifshitz1991} Here we denoted the normalization constant is
$N_{nm}=\left[  n!/(n+|m|)!\right]  ^{\frac{1}{2}}2^{-|m|/2}l_{B}^{-1}$,
$L_{n}^{|m|}(x)$ is the associated Laguerre polynomials, and $l_{B}%
=\sqrt{\hbar/eB}$ is the magnetic length. The boundary condition gives the
degeneracy factor of each Landau level, $m_{\text{max}}=R^{2}/2l_{B}^{2}$,
where $R$ is the fixed radius of the circular disk geometry (the system size).
The filling factor is defined as $\nu=N_{e}/m_{\text{max}}$, where $N_{e}$ is
the total number of electrons.

In this section, we provide the quantitative analysis of the induced
photocurrent via the transfer of optical AM. To see this, we investigate the
interaction between a Landau-quantized 2DEG and CPBOV at the $z=0$ plane. We
start with the total Hamiltonian, which is configured by the 2DEG Hamiltonian
(\ref{Non-perturmativeHamil}) and light-electron coupling one, $H_{\text{int}%
}=-\mathbf{A}_{\ell,\sigma}^{\text{OV}}\left(  \theta\right)  \cdot\mathbf{j}%
$:
\begin{equation}
H\mathbf{=}H_{0}+H_{\text{int}}=\frac{1}{2m_{e}^{\ast}}\left[  -i\hbar
\mathbf{\nabla}+e\mathbf{A}^{\text{ext}}(\mathbf{r})\right]  ^{2}%
-\mathbf{A}_{\ell,\sigma}^{\text{OV}}\left(  \theta\right)  \cdot\mathbf{j},
\end{equation}
where $\mathbf{A}_{\ell,\sigma}^{\text{OV}}\left(  \theta\right)  $ is the
spatial part of CPBOV vector potential which travels along the $Z$-axis
(\ref{Tilted_vector_potential_lab2}), and the photocurrent operator is given
by $\mathbf{j}=e(\mathbf{p}+e\mathbf{A}^{\text{ext}})/m_{e}$. By an extremely
strong magnetic field, the spin degree of freedom is frozen, and we can
neglect the degree of spin of an electron.

To see the induced photocurrent, we apply the linear response theory. This is
justified when the electric and magnetic fields of OV are sufficiently weak.
Thus we start with the Kubo formula for $i$ component of the response
current\cite{Kubo1957,Ando1975} :%
\begin{align}
\delta j_{i}(\omega)  & =-\frac{1}{V}\sum_{n,m}\sum_{n^{\prime},m^{\prime}%
}\left(  f(E_{n,m})-f(E_{n^{\prime},m^{\prime}})\right) \nonumber\\
& \times\frac{\langle n,m|j_{i}|n^{\prime},m^{\prime}\rangle\langle n^{\prime
},m^{\prime}|\mathbf{A}_{\ell,\sigma}^{\text{OV}}\left(  \theta\right)
\cdot\mathbf{j}|n,m\rangle}{E_{n,m}-E_{n^{\prime},m^{\prime}}+\hbar
\omega+i\delta}.\label{Kubo_formula1}%
\end{align}
where $V$ is the volume of the system and $f(E_{n,m})$ is the Fermi
distribution, $f(\epsilon)=\left[  \exp\beta\left(  \epsilon-\mu\right)
+1\right]  ^{-1}$ with the chemical potential $\mu$ and the inverse
temperature $\beta$. We note that we do not take account of any impurity
effects in this paper. Although the impurity-induced relaxation may provide
the imaginary part of the electron self-energy, $\tau^{-1}$, in a real system,
we added a positive infinitesimal $\delta$ in the denominator to give\ the
numerical analysis below. We concentrate on zero temperature, $T=0$.

The matrix element of the photocurrent operator in a chiral basis, $j_{\pm
}=\left(  j_{x}\mp ij_{y}\right)  /\sqrt{2}$, where the upper (under) sign
corresponds to the right (left)-circularly polarized photocurrent\ is written
as%
\begin{equation}
\langle n,m|j_{\pm}|n^{\prime},m^{\prime}\rangle=i\frac{e}{\sqrt{2}\hbar
}(E_{n,m}-E_{n^{\prime},m^{\prime}})C_{n,m}^{n^{\prime},m^{\prime}}%
\delta_{\Delta m,\pm1}%
\end{equation}
where we denoted $\Delta m=m^{\prime}-m$, and the radial integral as%
\begin{equation}
C_{n,m}^{n^{\prime},m^{\prime}}=\int d\rho\text{ }\rho^{2}\mathfrak{R}%
_{n^{\prime},m^{\prime}}(\rho)\mathfrak{R}_{n,m}(\rho).
\end{equation}
From the azimuthal integral, we obtain the selection rule,%
\begin{equation}
\Delta m=\pm1,\label{selectionrule1}%
\end{equation}
The +(-) in (\ref{selectionrule1})\ corresponds to the result of the right
(left)-circularly polarized current, respectively. Further by performing the
radial integral and calculating the energy factor, $E_{n,m}-E_{n^{\prime
},m^{\prime}}$ with Eq.(\ref{Energy_Landau}), we also obtain the matrix
elements for the photocurrent operator matrix as%
\begin{align}
\langle n,m|j_{+}|n^{\prime},m+1\rangle &  =%
\begin{cases}
-iel_{B}\omega_{c}\sqrt{n+\frac{\left\vert m\right\vert +m}{2}+1}, & \text{for
}n^{\prime}=n\text{ and }m\geq0,\\
iel_{B}\omega_{c}\sqrt{n+\frac{\left\vert m\right\vert +m}{2}+1}, & \text{for
}n^{\prime}=n+1\text{ and }m\leq-1,\\
0, & \text{for otherwise,}%
\end{cases}
\label{right-handed_matrix}\\
\langle n,m|j_{-}|n^{\prime},m-1\rangle &  =%
\begin{cases}
iel_{B}\omega_{c}\sqrt{n+\frac{\left\vert m\right\vert +m}{2}}, & \text{for
}n^{\prime}=n\text{ and }m\geq1,\\
-iel_{B}\omega_{c}\sqrt{n+\frac{\left\vert m\right\vert +m}{2}}, & \text{for
}n^{\prime}=n-1\text{ and }m\leq0,\\
0, & \text{for otherwise.}%
\end{cases}
\label{left-handed_matrix}%
\end{align}
Then, if the LLL, $N=0$, is filled by electrons (which corresponds to the
filling factor being set to $\nu=1$), the Pauli exclusion principle allows for
the transitions from the LLL to the 2LL, $N=1$. Consequently, the element,
$\langle n,m|j_{+}|n^{\prime},m^{\prime}\rangle$ survives only for
\begin{equation}
\Delta m=1.\label{selectionrule2a}%
\end{equation}
and the possible transitions are limited to the cases,
\begin{equation}%
\begin{array}
[c]{cccc}%
\left(  n,m,N\right)  &  & \left(  n^{\prime},m^{\prime},N^{\prime}\right)  &
\\
(0,0,0) & \rightarrow & (0,1,1), & \text{for }m=0,\\
(0,m,0) & \rightarrow & (1,m+1,1), & \text{for }m<0.
\end{array}
\label{selectionrule2b}%
\end{equation}

Next, to calculate the matrix element of the light-electron coupling, $\langle
n^{\prime},m^{\prime}|\mathbf{A}_{\ell,\sigma}^{\text{OV}}\left(
\theta\right)  \cdot\mathbf{j}|n,m\rangle$, we apply the long-wavelength
approximation. Then the matrix element reduces to%
\begin{align}
& \langle n^{\prime},m^{\prime}|\mathbf{A}_{\ell,\sigma}^{\text{OV}}\left(
\theta\right)  \cdot\mathbf{j}|n,m\rangle\nonumber\\
& \sim i\frac{e}{\hbar}\left(  E_{n^{\prime},m^{\prime}}-E_{n,m}\right)
\langle n^{\prime},m^{\prime}|\mathbf{A}_{\ell,\sigma}^{\text{OV}}\left(
\theta\right)  \cdot\mathbf{r}|n,m\rangle.\label{ElementInLWA}%
\end{align}
This matrix element depends on the incident angle $\theta$ via
(\ref{Tilted_vector_potential_lab2}). In this paper, we restrict our
discussion to the small incident angle case, $\theta\ll1$. After some
calculations, we obtain the matrix element with $\theta$-dependence,%
\begin{align}
& \left\langle n^{\prime},m^{\prime}\left\vert \mathbf{A}_{\ell,\sigma
}^{\text{OV}}\left(  \theta\right)  \cdot\mathbf{r}\right\vert
n,m\right\rangle \nonumber\\
& =A_{0}\sqrt{\frac{k_{\perp}}{4\pi}}{\sum\limits_{L=-\infty}^{\infty}%
}\mathrm{e}^{i\frac{\pi}{2}\left(  L+1\right)  }\delta_{\Delta m,\ell
+\sigma+L}\nonumber\\
& \times\int d\rho\text{ }\rho^{2}\mathfrak{R}\left(  \rho\right)
\mathfrak{R}_{n_{,}m,}\left(  \rho\right)  J_{\ell}\left(  k_{\perp}%
\rho\right)  J_{L}\left(  k_{\parallel}\rho\text{ }\theta\right)  .
\end{align}
The azimuthal angular integral gives the AM conservation,%
\begin{equation}
\Delta m=\ell+\sigma+L.
\end{equation}
By combining this conservation, optical SAM, $\sigma=\pm1$, and Eq.
(\ref{selectionrule2a}), it gives the modified AM selection rules,
\begin{equation}
\ell+L=0,\sigma=1,\text{or }\ell+L=2,\sigma=-1,
\end{equation}
which is different from the vertically irradiated case (\ref{Selection_rule0}%
). As $L$ can be any integer number, the initial optical OAM $\ell$\ can also
be allowed to be an integer number. In other words, it is allowed that the
2DEG absorbs the OV beam with any integer OAM $\ell$\ in oblique incidence.
Putting $\theta=0$ in (\ref{Tilted_vector_potential_lab2}), only the
$L=0$\ term survives and the result of the vertical incidence is reproduced.

The summation over $L$ leads to%
\begin{equation}
\left\langle n^{\prime},m^{\prime}\left\vert \mathbf{A}_{\ell,\sigma
}^{\text{OV}}\left(  \theta\right)  \cdot\mathbf{r}\right\vert
n,m\right\rangle =\mathrm{e}^{i\frac{\pi}{2}\left(  \Delta m-\ell
-\sigma+1\right)  }A_{0}\sqrt{\frac{k_{\perp}}{4\pi}}D_{n,m,\ell}^{n^{\prime
},m^{\prime}}\left(  \theta\right)  ,
\end{equation}
where we denoted the radial integral in the above expression as
\begin{equation}
D_{n,m,\ell,\sigma}^{n^{\prime},m^{\prime}}\left(  \theta\right)  =\int
d\rho\text{ }\rho^{2}\mathfrak{R}_{n^{\prime},m^{\prime}}\left(  \rho\right)
\mathfrak{R}_{n,m}\left(  \rho\right)  J_{\ell}\left(  k_{\perp}\rho\right)
J_{\Delta m-\ell-\sigma}\left(  k_{\parallel}\rho\text{ }\theta\right)  .
\end{equation}

As we obtained the expressions of matrix elements, $\langle n,m|j_{i}%
|n^{\prime},m^{\prime}\rangle$, and\ $\langle n^{\prime},m^{\prime}%
|\mathbf{A}_{\ell,\sigma}^{\text{OV}}\left(  \theta\right)  \cdot
\mathbf{j}|n,m\rangle$, we now turn to the photocurrent using the Kubo formula
(\ref{Kubo_formula1}). We now concentrate our discussion on zero temperature
and the filling factor is set to $\nu=1$. Then, by (\ref{selectionrule2b}),
the left-circularly polarized current is not induced and the right one just
arises in transitions from $N=0$ to $N=1$. The $\theta$-dependent OV-induced
photocurrent then reduces to%
\begin{equation}
\delta j_{\ell,\sigma}^{+}\left(  \theta,\omega,B\right)  =-\mathrm{e}%
^{i\frac{\pi}{2}\left(  2-\ell-\sigma\right)  }\frac{F_{\sigma}^{\ell}\left(
\theta,B\right)  }{\hbar\omega-\hbar\omega_{c}+i\delta},\label{Kubo_formula2}%
\end{equation}
and the factors $F_{\sigma}^{\ell}\left(  \theta,B\right)  $ are given by
\begin{equation}
F_{\sigma}^{\ell}\left(  \theta,B\right)  =A_{1}C_{0,0}^{0,1}D_{0,0,\ell
}^{0,1}\left(  \theta\right)  +A_{1}\sum_{m<0}^{-m_{\max}}C_{0,m}%
^{1,m+1}D_{0,m,\ell,\sigma}^{1,m+1}\left(  \theta\right)  ,\label{I02}%
\end{equation}
with $A_{1}=A_{0}e^{2}\omega_{c}^{2}\sqrt{k_{\perp}/2\pi}/V$. In the summation
over $m$, only the term corresponding to the edge state survives and the other
terms corresponding to the bulk states cancel each other. That is, the induced
photocurrent localizes on the system edge $R$.

To see the net contribution of $\delta j_{\ell,\sigma}^{+}\left(
\theta,\omega,B\right)  $, we evaluate the energy denominator. When we take a
clean limit $\delta\rightarrow0$, the energy denominator is decomposed into
two terms as%
\begin{equation}
\frac{1}{\hbar\omega-\hbar\omega_{c}+i\delta}\rightarrow\mathcal{P}\frac
{1}{\hbar\omega-\hbar\omega_{c}}-i\delta(\hbar\omega-\hbar\omega_{c}),
\end{equation}
where $\mathcal{P}$\ indicates Cauchy principal value. We assume this
principal value can be neglected around $\omega\sim\omega_{c}$.\ Then
$\delta(\hbar\omega-\hbar\omega_{c})$ term gives the energy conservation and a
net contribution to the induced photocurrent. As the cyclotron frequency
$\omega_{c}$\ depends on the magnetic field $B$,\ the optical frequency
$\omega$ also is linked to the magnetic field $B$ via $\delta(\hbar
\omega-\hbar\omega_{c})$.\ Therefore, to observe the $B$ dependence of $\delta
j_{\ell,\sigma}^{+}\left(  \theta,\omega_{c},B\right)  $, it\ is needed
the\ change\ of the optical frequency $\omega$\ or\ the magnitude of the
wavenumber $k$\ to keep $k=\omega_{c}/c$.

By incorporating magnetic field dependence and using\ $\alpha\ll1$ (paraxial
approximation), we approximately obtain the factor $F_{\sigma}^{\ell}\left(
\theta,B\right)  $ for the OV beam as%
\begin{align}
F_{\sigma}^{\ell}\left(  \theta,B\right)   &  =F_{0}\left(  \frac{1+\alpha
^{2}}{\alpha^{2}}\frac{\Phi_{0}^{2}}{\lambda_{e}^{2}B^{2}R^{2}}\left[
1+\frac{\Phi_{0}}{2\pi BR^{2}}\right]  \mathrm{e}\right)  ^{\pi R^{2}%
B/\Phi_{0}}\nonumber\\
&  \times\int dx\text{ }x^{2m_{\max}+3}\exp\left(  -\frac{x^{2}}{2k_{\perp
}l_{B}}\right)  J_{\ell}\left(  x\right)  J_{1-\ell-\sigma}\left(  \frac
{x}{\alpha}\text{ }\theta\right)  ,\label{Final expression}%
\end{align}
where $F_{0}=A_{0}e^{2}c^{2}/V\sqrt{4\pi\lambda_{e}\mathrm{e}}$, $\Phi
_{0}\left(  =2\pi\hbar/e\right)  $ is the flux quantum, $\lambda_{e}\left(
=2\pi\hbar/m_{e}c\right)  $ is the electron Compton wavelength, and
$\alpha\left(  =k_{\perp}/k_{\parallel}\right)  $ is the ratio of paraxial
approximation. Here, $x\left(  =k_{\perp}\rho\right)  $ is a dimensionless variable.

Next, we grasp the overall profile of these qualitative behaviors through the
integral in Eq. (\ref{Final expression}). We decompose the integrand into
\begin{equation}
g(x)=x^{2m_{\text{max}}+3}\mathrm{\exp}\left(  -\frac{x^{2}}{2k_{\perp}%
^{2}l_{B}^{2}}\right)
\end{equation}
and the Bessel functions,
\begin{equation}
h\left(  x,\theta\right)  =J_{\ell}\left(  x\right)  J_{1-\ell-\sigma}\left(
\frac{x\theta}{\alpha}\right)  .
\end{equation}
We note that $g(x)$ has its extrema at
\begin{equation}
x^{\ast}=\sqrt{2m_{\text{max}}+3}k_{\perp}l_{B}\simeq587\alpha RB,
\end{equation}
corresponds to the system edge $R$. Using the values $\alpha=0.1$, $R=10^{-2}$
m, $m_{e}=9.11\times10^{-31}$ kg, and $c=3.00\times10^{8}$ m/s,\ we have
$x^{\ast}\simeq0.587B$ with $B$ being measured in Tesla. On the other hand,
the Bessel functions factor $h\left(  x,\theta\right)  $\ describes the
interference between the initial CPBOV beam profile $J_{\ell}\left(  x\right)
$\ and $J_{1-\ell-\sigma}\left(  x\theta/\alpha\right)  $. Because of the
positive roots of the Bessel functions, $J_{\ell}\left(  x\right)  =0$\ and
$J_{1-\ell-\sigma}\left(  x\theta/\alpha\right)  =0$, $h\left(  x,\theta
\right)  $ manifests\ zero lines\ in $x\theta$-plane and demonstrate
oscillating behavior. Consequently, $g(x)$ and $h\left(  x,\theta\right)  $
significantly interfere with each other and behave oscillation. Thus when the
zeros of $J_{\ell}\left(  x\right)  =0$\ or $J_{1-\ell-\sigma}\left(
x\theta/\alpha\right)  =0$,\ overlaps the peak\ of $g(x)$, namely $x^{\ast}$,
$F_{\sigma}^{\ell}\left(  \theta,B\right)  $ is disappeared. The physical
meaning of this is that the OV-induced photocurrents are disappeared when the
dark rings constituted through the zeros of $J_{\ell}\left(  k_{\perp}%
\rho\right)  =0$\ and$\ J_{1-\ell-\sigma}\left(  k_{\parallel}\rho\text{
}\theta\right)  =0$\ overlap the system edge $R$.

Here, to elucidate the salient feature of OV, it may be useful to consider the
case of a purely plane wave. To see this, we give the expression of the
induced photocurrent by absorption of the circularly polarized plane wave
(CPPW). The vector potential of the CPPW traveling along the $Z$-axis is given
by%
\begin{equation}
\boldsymbol{A}_{\sigma}^{\text{PW}}(\boldsymbol{r}|\theta)=iA_{0}^{\text{PW}%
}\mathbf{\varepsilon}_{\sigma}\mathrm{e}^{ik_{\parallel}Z}%
,\label{vector_potential_PW}%
\end{equation}
where $A_{0}^{\text{PW}}$ is a real constant and $\mathbf{\varepsilon}%
_{\sigma}$\ is the circular polarization in the beam flame (\ref{Cir_pol}). As
given in Appendix \ref{Current_CPPW}, the PW-induced photocurrent for
$\theta\ll1$ reads%
\begin{equation}
\delta j_{+}^{\text{PW}(\sigma)}(\omega)=-i\sigma\frac{F_{\sigma}^{\text{PW}%
}\left(  \theta,B\right)  }{\hbar\omega-\hbar\omega_{c}+i\delta}%
,\label{Kubo_formulaPW}%
\end{equation}
where the factor $F_{\sigma}^{\text{PW}}\left(  \theta,B\right)  $ is given by%
\begin{align}
F_{\sigma}^{\text{PW}}\left(  \theta,B\right)   &  =F_{0}^{\text{PW}}%
B^{5/2}\left(  \frac{\mathrm{e}}{R^{2}}\left[  1+\frac{\Phi_{0}}{2\pi R^{2}%
B}\right]  \right)  ^{\frac{\pi R^{2}}{\Phi_{0}}B}\nonumber\\
&  \times\int d\rho\text{ }\rho^{2m_{\max}+3}\exp\left(  -\frac{\rho^{2}%
}{2l_{B}^{2}}\right)  J_{1-\sigma}(k_{\parallel}\rho\ \theta
),\label{Final expressionPW}%
\end{align}
with $F_{0}^{\text{PW}}=A_{0}^{\text{PW}}e^{2}c^{2}\lambda_{e}^{2}%
/V\sqrt{2\Phi_{0}^{5}\mathrm{e}}$. Each helicity wave can induce
photocurrents. In other words, that the AM selection rule for CPPW absorption
is $\sigma=\pm1 $.\ Unlike $J_{\ell}\left(  k_{\perp}\rho\right)
J_{1-\ell-\sigma}\left(  k_{\parallel}\rho\text{ }\theta\right)  $\ in
$F_{\sigma}^{\ell}\left(  \theta,B\right)  $, $J_{1-\sigma}(k_{\parallel}%
\rho\ \theta)$ appears in $F_{\sigma}^{\text{PW}}\left(  \theta,B\right)  $.
The disappearance\ of PW-induced photocurrent arises from an overlap of only
the positive roots of $J_{1-\sigma}(k_{\parallel}\rho\ \theta)=0.$\ Also, when
$\theta=0$, it gives $F_{-1}^{\text{PW}}\left(  \theta,B\right)  =0$, and
$F_{1}^{\text{PW}}\left(  \theta,B\right)  \neq0$. Thus, when the vertical
incidence, only the CPPW with positive helicity, $\sigma=1$, induces the
photocurrent. As the CPPW carries no OAM, this corresponds to one of the
OV-absorption selection rules, $\ell=0$, $\sigma=1$.

\section{Numerical Results}

\begin{figure}
[ptb]
\begin{center}
\includegraphics[
height=7.0716in,
width=6.2673in
]%
{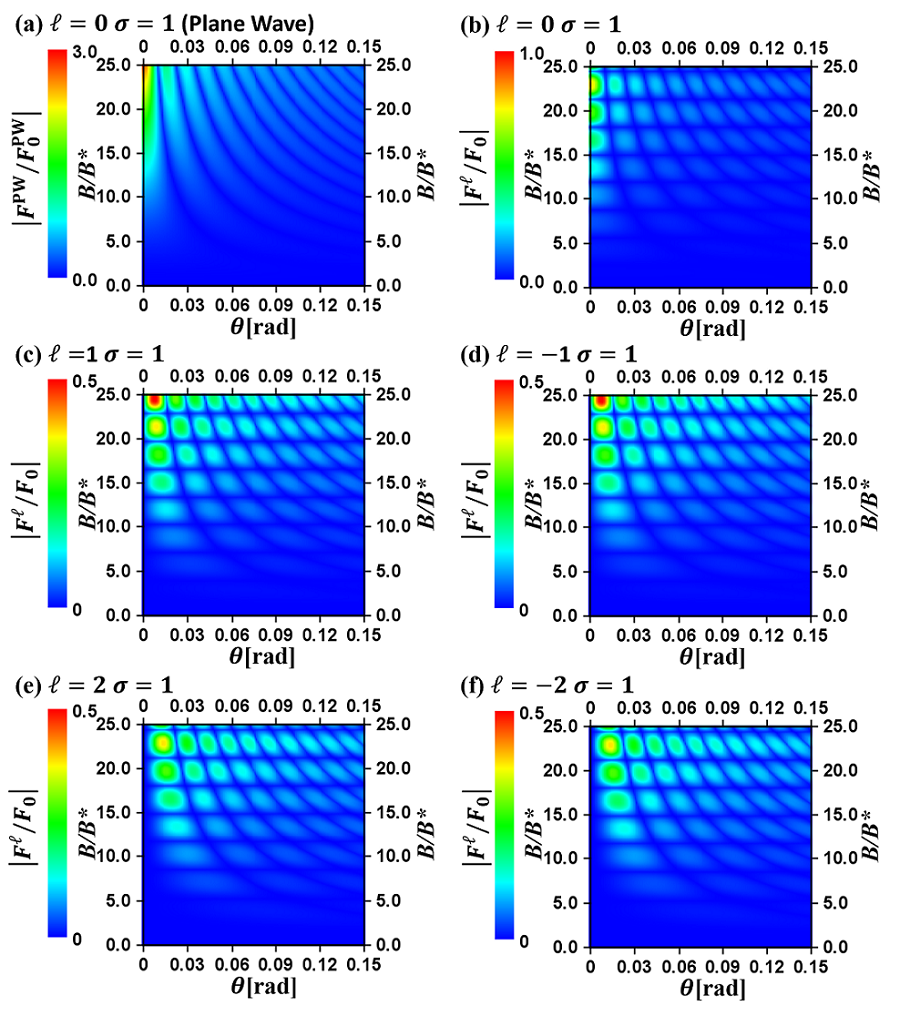}%
\caption{The contour plots of the $\theta$ and $B$ dependence of intensity
$F_{\sigma}^{\ell} $ and $F_{\sigma}^{\text{PW}}$\ for positive helicity waves
when the chemical potential is kept between the LLL and second LL. (a) plane
wave ($\ell=0$), (b) vortex wave with $\ell=0$, (c) $\ell=1$, (d) $\ell=-1$,
(e) $\ell=2$, (f) $\ell=-2$. Parameters are $R=10^{-2}$~m in system size and
$\alpha=0.1$ with $\alpha=k_{\perp}/k_{\parallel}$. The wavenumber has $B$
dependence as $k=5.87\times10^{2}B$ [m$^{-1}$].\ The vertical axes are scaled
by $B^{\ast}=\Phi_{0}/\alpha\lambda_{e}R=1.70\times10^{-3}/\alpha R$ [T].}%
\label{Num_Result1}%
\end{center}
\end{figure}

\begin{figure}
[ptb]
\begin{center}
\includegraphics[,
height=7.0716in,
width=6.2673in
]%
{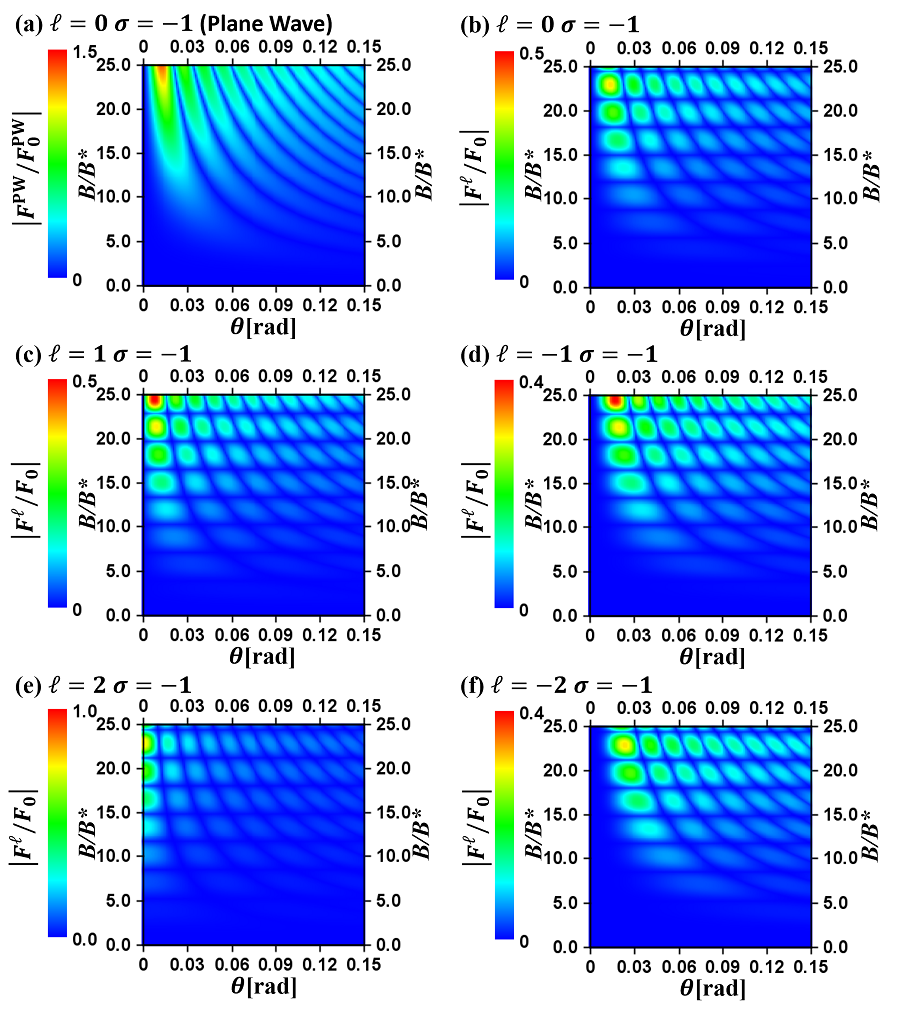}%
\caption{The contour plots of the $\theta$ and $B$ dependence of intensity
$F_{\sigma}^{\ell}$ and $F_{\sigma}^{\text{PW}}$\ for negative helicity waves
when the chemical potential is kept between the LLL and second LL. (a) plane
wave ($\ell=0$), (b) vortex wave with $\ell=0$, (c) $\ell=1$, (d) $\ell=-1$,
(e) $\ell=2$, (f) $\ell=-2$. Parameters are $R=10^{-2}$~m in system size and
$\alpha=0.1$ with $\alpha=k_{\perp}/k_{\parallel}$. The wavenumber has $B$
dependence as $k=5.87\times10^{2}B$ [m$^{-1}$].\ The vertical axes are scaled
by $B^{\ast}=\Phi_{0}/\alpha\lambda_{e}R=1.70\times10^{-3}/\alpha R$ [T].}%
\label{Num_Result2}%
\end{center}
\end{figure}

In this section, we demonstrate the numerical results of the incident angle
$\theta$ and magnetic field $B$\ dependence of $\delta j_{\ell,\sigma}%
^{+}\left(  \theta,\omega,B\right)  $.\ Figure \ref{Num_Result1} and
\ref{Num_Result2} show the magnetic field $B$ and incident angle $\theta$
dependencies of $F_{\sigma}^{\ell}\left(  \theta,B\right)  $ for $\ell
=0,\pm1,\pm2$\ with each helicity when we choose the parameters as
$\alpha=0.1$, $R=10^{-2}$ m, $m_{e}=9.11\times10^{-31}$ kg, and $c=3.00\times
10^{8}$ m/s. Also, we give the results of the plane wave case, $F_{\sigma
}^{\text{PW}}\left(  \theta,B\right)  $, in these figures. We here introduce
the characteristic magnetic field strength, $B^{\ast}\equiv\Phi_{0}%
/\alpha\lambda_{e}R$, which corresponds to $k_{\perp}R\sim1.$ Note that the
wavelengths (wavenumbers) are tuned with the change of the applied magnetic
field $B$ to induce the photocurrent via energy conservation.%


It is notable that the absorption amplitudes $\left\vert F_{\sigma}^{\ell
}\left(  \theta,B\right)  \right\vert $\ oscillate and evolve with increasing
the magnetic field strength $B$. The origin of the oscillating behavior in
change $B$ is as follows.\ When $B$ changes, the dark ring radius of the OV
intensity which is determined by the positive root of $J_{\ell}\left(
k_{\perp}\rho\right)  =0$ changes with the transverse wavenumber $k_{\perp
}\left(  =587\alpha B\right)  $\ with fixed $\alpha$.\ When the dark ring
overlaps the system edge $R$\ where the photocurrent is localizing, the
induced photocurrent is disappeared.\ Also, the origin of this evolutional
behavior in change $B$ is as follows.\ When $B$ increases, it narrows the
width of the electron wavefunction which is linked by the magnetic length
$l_{B}$. Then the number of electrons in the LLL becomes increasing because
the chemical potential is kept between the LLL ($N=0$) and the second LL
($N=1$). Thus the number of electrons contributing to the induced photocurrent
increases in increasing $B$. Further, it is noted that the absorption
amplitudes oscillate in the change of the incident angle $\theta$. The reason
is why the dark rings of obliquely irradiated OV intensity depend on the
positive roots of $J_{1-\ell-\sigma}(k_{\parallel}\rho\ \theta)=0$ as well.
When the dark ring in the change $\theta$\ overlaps the system edge $R$, the
induced photocurrent is disappeared in analogy with the case of change $B$.

Consequently, there exist the lattice-shaped disappearance pattern due to the
interference between the positive roots of Bessel function, $J_{\ell}%
(k_{\perp}\rho)=0$, and $J_{1-\ell-\sigma}\left(  k_{\parallel}\rho
\ \theta\right)  =0$.\ In particular, the vertical incidence, $\theta=0$,
gives no induced photocurrent except for $\ell=0$, $\sigma=1,\ $and $\ell=2$,
$\sigma=-1$. These results are equivalent to the results shown in our previous
paper.\cite{Takahashi2018,Takahashi2019}

\section{Conclusions and Remarks}

We have investigated the modification of selection rules in induced
photocurrents by coupling the Landau-quantized system with obliquely
irradiated OV beams. As discussed in our previous paper, the Landau-quantized
system is suitable for causing interaction with OVs. The reasons are as
follows: It is known that the optical OAM can not be transferred to\ the
internal motion but the center-of-mass motion in an electric dipole
transition.\cite{Babiker2002} To transfer the optical OAM to\ the electron
center-of-mass motion through the electric dipole transition, it is needed to
match the OV spatial structure with the electron system geometry. In other
words, if the intensity profile of OV beams has an axially symmetrical
structure around the optical axis, the electron system geometry should have
axial symmetry. Therefore, we considered that the best choice is a 2DEG
confined on a circular disk geometry. Furthermore, to discuss the absorption
of the orbital and spin AM, a paraxial approximation of OV beams is important.
Absorptions of optical AM to the photocurrent are confined to optical TAM of
one, $J=1$. Thus, in the vertical incidence, the combinations of the optical
OAM and SAM are locked as only Eq. (\ref{Selection_rule0}) because of the
conservation of OAM.\cite{Takahashi2018} However, it is inevasible to consider
the case of oblique incidence from the experimental viewpoints. Since the
obliquely irradiated (angular deflected) beam consists of a superposition of
the various OAM, $\ell+L$, with initial optical OAM $\ell$ and arbitrary orbit
number $L$.\cite{Vasnetsov2005} Thus, in this paper, it was explicitly shown
that the partial wave carrying the OAM, $\ell+L$, is absorbed by 2DEG and it
modifies AM selection rule, $\ell+\sigma+L=1$. In other words, absorptions of
any integer initial optical OAM $\ell$ are allowed in the oblique incidence.

We also showed oscillating behaviors of the OV-induced photocurrent\ in a
change of the incident angle $\theta$ and the magnetic field $B$. The
oscillating behavior has its origin in the interferences between the profile
of light intensity and that of the current localized on the system edge.
Especially, as the obliquely irradiated OV beam seems a superposition of
partial waves with the $\theta$-dependent\ Bessel function type
amplitude,\cite{Vasnetsov2005} the OV-induced current also has $\theta
$-dependent oscillating behavior. On the other hand, it is noted that as the
obliquely irradiated purely plane wave is also described by a superposition of
partial wave with the $\theta$-dependent Bessel function type amplitude, the
$\theta$-dependent oscillating behavior is yielded even in the case of the
purely plane wave. Nevertheless, as the OV has a characteristic intensity
profile (Bessel-mode, LG-mode, and elsewhere), the oscillating behavior
different from the case of the purely plane wave is yielded.

In our numerical results, the bare electron mass was used. However, the
effective (cyclotron) mass $m_{e}^{\ast}$\ should be used for actual
materials. As the cyclotron frequency, $\omega_{c}=eB/m_{e}^{\ast}$, is then
changed, the wavenumber is also changed via the energy conservation in
transitions from the LLL to the second LL. This affects the radius of the
bright and dark rings. For example, for the widely accepted value,
$m_{e}^{\ast}=0.067m_{e}$, for GaAs, the wavenumber of the OV beam is
evaluated as $k/0.067$. Then the bright and dark rings become $0.067$ times
larger in radius. Therefore, it is needed to pay attention to suitable 2DEG
system size $R$ to match with the OV spatial structure in experiments.

\begin{acknowledgments}
The authors thank Dr. S. Hashiyada, I. Proskurin, Professors Y. Yamada, K.
Oto, and Y. Togawa for fruitful discussions. This work was supported by JSPS
KAKENHI Grant (25220803, 17H02923), and the JSPS Bilateral (Japan-Russia)
Joint Research Projects.
\end{acknowledgments}

\appendix

\section{Long-Wavelength Approximation in Light-Electron Coupling Hamiltonian}

In this Appendix, we discuss the validity of the long-wavelength approxmation
in Eq. (\ref{ElementInLWA}). We first show the commutation relation,
$[H_{0},\mathbf{A}^{\text{OV}}]$, with non-perturbative Hamiltonian,
$H_{0}=\frac{1}{2m_{e}^{\ast}}\left(  -i\hbar\mathbf{\nabla}+e\mathbf{A}%
^{\text{ext}}\right)  ^{2}$. Note that a vector potential of OV is described
by $\mathbf{A}^{\text{OV}}=\mathbf{\varepsilon}_{\sigma}f\left(  k_{\perp}%
\rho\right)  \mathrm{e}^{i\ell\phi+ik_{\parallel}z}$ with $\mathbf{\nabla
}\cdot\mathbf{A}^{\text{OV}}=0$, and $\operatorname{grad}\left[  f\left(
k_{\perp}\rho\right)  \mathrm{e}^{i\ell\phi+ik_{\parallel}z}\right]
\sim\mathbf{k}$ $f\left(  k_{\perp}\rho\right)  \mathrm{e}^{i\ell
\phi+ik_{\parallel}z}$. After some calculations, we obtain%
\begin{equation}
H_{0}\mathbf{A}^{\text{OV}}\approx\mathbf{A}^{\text{OV}}H_{0}+\mathbf{A}%
^{\text{OV}}\left\{  \frac{\hbar^{2}k^{2}}{2m_{e}^{\ast}}-\left(  \frac
{\hbar\mathbf{k}}{m_{e}^{\ast}}\right)  \cdot(-i\hbar\mathbf{\nabla})+\frac
{e}{m_{e}^{\ast}}(\mathbf{A}^{\text{ext}}\cdot\hbar\mathbf{k})\right\}  .
\end{equation}
Therefore, the commutation relation is written by%
\begin{equation}
\lbrack H_{0},\mathbf{A}^{\text{OV}}]=\mathbf{A}^{\text{OV}}\left\{
\frac{\hbar^{2}k^{2}}{2m_{e}^{\ast}}-\left(  \frac{\hbar\mathbf{k}}%
{m_{e}^{\ast}}\right)  \cdot(-i\hbar\mathbf{\nabla})+\frac{e}{m_{e}^{\ast}%
}(\mathbf{A}^{\text{ext}}\cdot\hbar\mathbf{k})\right\}  .\label{Comm_Rel}%
\end{equation}

Next, we evaluate the matrix element of the light-electron coupling, $\langle
n^{\prime},m^{\prime}|\mathbf{A}^{\text{OV}}\cdot\mathbf{j}|n,m\rangle$, with
a photocurrent operator, $\mathbf{j}=\frac{e}{m_{e}^{\ast}}\left(
\mathbf{p}+e\mathbf{A}^{\text{ext}}\right)  $. Noting that the photocurrent
operator satisfies
\begin{equation}
\mathbf{j=p}+e\mathbf{A}^{\text{ext}}=\frac{im_{e}^{\ast}}{\hbar}%
[H_{0},\mathbf{r}]
\end{equation}
by using the commutation relation (\ref{Comm_Rel}), we obtain
\begin{align}
\langle n^{\prime},m^{\prime}|\mathbf{A}^{\text{OV}}\cdot\mathbf{j}|n,m\rangle
&  =\frac{e}{m_{e}}\langle n^{\prime},m^{\prime}|\mathbf{A}^{\text{OV}}%
\cdot\frac{im_{e}}{\hbar}[H_{0},\mathbf{r}]|n,m\rangle\nonumber\\
&  =\frac{ie}{\hbar}(E_{n^{\prime},m^{\prime}}-E_{n,m})\langle n^{\prime
},m^{\prime}|\mathbf{A}^{\text{OV}}\cdot\mathbf{r}|n,m\rangle\nonumber\\
&  -\frac{ie}{\hbar}\frac{\hbar^{2}k^{2}}{2m_{e}}\langle n^{\prime},m^{\prime
}|\mathbf{A}^{\text{OV}}\cdot\mathbf{r}|n,m\rangle\nonumber\\
&  -\frac{ie}{\hbar}\frac{\hbar e}{m_{e}}\langle n^{\prime},m^{\prime
}|(\mathbf{A}^{\text{ext}}\cdot\mathbf{k})(\mathbf{A}^{\text{OV}}%
\cdot\mathbf{r})|n,m\rangle\nonumber\\
&  -\frac{e}{\hbar}\frac{\hbar^{2}}{2m_{e}}\langle n^{\prime},m^{\prime
}|\mathbf{A}^{\text{OV}}\cdot\mathbf{k}|n,m\rangle\nonumber\\
&  -\frac{e}{\hbar}\frac{\hbar^{2}}{2m_{e}}\langle n^{\prime},m^{\prime
}|(\mathbf{A}^{\text{OV}}\cdot\mathbf{r})\mathbf{k}\cdot\operatorname{grad}%
|n,m\rangle.\label{Minimal_coupling_transition}%
\end{align}
In the long-wavelength approximation, the second term is the second order in
$k$. It is $10^{9}$ times weaker than the first term in $B=10$ T and can be
dropped. Furthermore, as the third, fourth, and fifth terms in
(\ref{Minimal_coupling_transition}) have the first order in $k$, which is
$10^{5}$ times weaker than the first term in $B=10$ T. Then these terms can
also be ignored by the comparison with the first term. As a consequence, the
first term only survives in the matrix element of the light-electron coupling,%
\begin{equation}
\langle n^{\prime},m^{\prime}|\mathbf{A}^{\text{OV}}\cdot\mathbf{j}%
|n,m\rangle\sim\frac{ie}{\hbar}(E_{n^{\prime},m^{\prime}}-E_{n,m})\langle
n^{\prime},m^{\prime}|\mathbf{A}^{\text{OV}}\cdot\mathbf{r}|n,m\rangle.
\end{equation}
which is Eq. (\ref{ElementInLWA}).

\section{Induced Current by Circularly Polarized Plane
Wave\label{Current_CPPW}}

To compare the OV-induced photocurrent with PW-induced one, we derive
(\ref{Kubo_formulaPW}) and (\ref{Final expressionPW}). In the case of the
circularly polarized plane wave (\ref{vector_potential_PW}), its vector
potential in the Laboratory flame is given by%
\begin{align}
\boldsymbol{A}_{\sigma}^{\text{PW}}(\boldsymbol{r}|\theta) &  =\frac
{A_{0}^{\text{PW}}}{\sqrt{2}}(\boldsymbol{e}_{x}\cos\theta+i\sigma
\boldsymbol{e}_{y}-\boldsymbol{e}_{z}\sin\theta)\nonumber\\
&  \times{\sum\limits_{L=-\infty}^{\infty}}J_{L}(k_{\parallel}\rho\sin
\theta)\mathrm{e}^{i\frac{\pi}{2}\left(  L+1\right)  }\mathrm{e}^{iL\phi
}\mathrm{e}^{ik_{\parallel}z\cos\theta}.
\end{align}
Then the matrix element of the light-electron coupling is given by%
\begin{align}
&  \left\langle n^{\prime},m^{\prime}\left\vert \boldsymbol{A}_{\sigma
}^{\text{PW}}\left(  \theta\right)  \cdot\boldsymbol{r}\right\vert
n,m\right\rangle \nonumber\\
&  \sim\frac{A_{0}^{\text{PW}}}{\sqrt{2}}{\sum\limits_{L=-\infty}^{\infty}%
}\mathrm{e}^{\frac{\pi}{2}\left(  L+1\right)  }\int d\rho\text{ }\rho
^{2}R_{n^{\prime},m^{\prime}}\left(  \rho\right)  R_{n,m}\left(  \rho\right)
J_{L}(k_{\parallel}\rho\text{ sin}\theta)\nonumber\\
&  \times\left\{  \cos^{2}\frac{\theta}{2}\delta_{\Delta m,L+\sigma}-\sin
^{2}\frac{\theta}{2}\delta_{\Delta m,L-\sigma}\right\}  .
\end{align}
When we assume $\theta\ll1$, we obtain%
\begin{align}
&  \left\langle n^{\prime},m^{\prime}\left\vert \boldsymbol{A}_{\sigma
}^{\text{PW}}\left(  \theta\right)  \cdot\boldsymbol{r}\right\vert
n,m\right\rangle \nonumber\\
&  \sim\frac{A_{0}^{\text{PW}}}{\sqrt{2}}{\sum\limits_{L=-\infty}^{\infty}%
}\mathrm{e}^{\frac{\pi}{2}\left(  L+1\right)  }\int d\rho\text{ }\rho
^{2}R_{n^{\prime},m^{\prime}}\left(  \rho\right)  R_{n,m}\left(  \rho\right)
J_{L}(k_{\parallel}\rho\text{ }\theta)\text{ }\delta_{\Delta m,L+\sigma},
\end{align}
where the second term was dropped because of $O\left(  \theta^{2}\right)  $.
After the summation over $L$, the matrix element reduces to%
\begin{align}
&  \left\langle n^{\prime}m^{\prime}\left\vert \boldsymbol{A}_{\sigma
}^{\text{PW}}\left(  \theta\right)  \cdot\boldsymbol{r}\right\vert
n,m\right\rangle \nonumber\\
&  \sim\frac{A_{0}^{\text{PW}}}{\sqrt{2}}\mathrm{e}^{\frac{\pi}{2}\left(
\Delta m-\sigma+1\right)  }\int d\rho\rho^{2}R_{n^{\prime},m^{\prime}}\left(
\rho\right)  R_{n,m}\left(  \rho\right)  J_{\Delta m-\sigma}(k_{\parallel}%
\rho\text{ }\theta).
\end{align}
When we proceed with the same procedure for the OV case to obtain the induced
photocurrent, it gives%
\begin{equation}
\delta j_{\sigma}^{+\text{PW}}(\theta,\omega,B)=-i\sigma\frac{F_{\sigma
}^{\text{PW}}\left(  B,\theta\right)  }{\hbar\omega-\hbar\omega_{c}+i\delta},
\end{equation}
where we defined the factor as%
\begin{align}
F_{\sigma}^{\text{PW}}\left(  B,\theta\right)   &  =F_{0}^{\text{PW}}%
B^{5/2}\left(  \frac{\mathrm{e}}{R^{2}}\left[  1+\frac{\Phi_{0}}{2\pi R^{2}%
B}\right]  \right)  ^{\frac{\pi R^{2}}{\Phi_{0}}B}\nonumber\\
&  \times\int_{0}^{R}d\rho\text{ }\rho^{2m_{\max}+3}\exp\left(  -\frac
{\rho^{2}}{2l_{B}^{2}}\right)  J_{-\sigma+1}(k_{\parallel}\rho\ \theta),
\end{align}
with $F_{0}^{\text{PW}}=A_{0}^{\text{PW}}e^{2}c^{2}\lambda_{e}^{2}%
/V\sqrt{2\Phi_{0}^{5}\mathrm{e}}.$ These are Eqs. (\ref{Kubo_formulaPW}) and
(\ref{Final expressionPW}).

\end{document}